\documentclass[journal]{vgtc}                
\ifpdf
  \pdfoutput=1\relax                   
  \usepackage{graphicx}                
  \DeclareGraphicsExtensions{.pdf,.png,.jpg,.jpeg} 
\else
  \usepackage{graphicx}                
  \DeclareGraphicsExtensions{.eps}     
\fi%

\graphicspath{{figures/}{pictures/}{images/}{./}} 

\usepackage{microtype}                 
\PassOptionsToPackage{warn}{textcomp}  
\usepackage{textcomp}                  
\usepackage{mathptmx}                  
\usepackage{times}                     
\usepackage{cite}                      
\usepackage{tabu}                      
\usepackage{booktabs}                  
\usepackage{xcolor}
\usepackage{inconsolata}
\usepackage{array}
\usepackage{stfloats}
\usepackage{wrapfig}

\newcolumntype{C}[1]{>{\centering\arraybackslash}m{#1}}

\definecolor{colorPOS}{HTML}{B257A0}
\definecolor{colorShape}{HTML}{4EA68B}
\definecolor{colorDimensionality}{HTML}{C2A11E}
\definecolor{colorScope}{HTML}{2B4BCA}
\definecolor{colorImagery}{HTML}{9B5C46}
\definecolor{colorEncoding}{HTML}{D44955}
\definecolor{colorView}{HTML}{307FB0}

\newcommand{\iconPOSInteractive}{\raisebox{-0.5ex}{\includegraphics[width=1em]{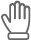}}}
\newcommand{\iconPOSAutomatic}{\raisebox{-0.5ex}{\includegraphics[width=1.4em]{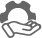}}}
\newcommand{\iconShapeFixed}{\raisebox{-0.5ex}{\includegraphics[width=1.2em]{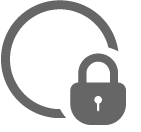}}}
\newcommand{\iconShapeDinamicUser}{\raisebox{-1.0ex}{\includegraphics[width=1.5em]{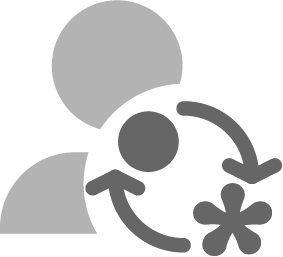}}}
\newcommand{\iconShapeDinamicData}{\raisebox{-1.0ex}{\includegraphics[width=1.6em]{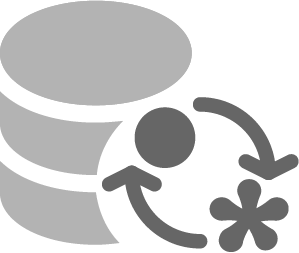}}}
\newcommand{\iconDimTwoD}{\raisebox{-0.5ex}{\includegraphics[width=1.2em]{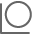}}}
\newcommand{\iconDimTwoHalfD}{\raisebox{-0.5ex}{\includegraphics[width=1.6em]{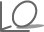}}}
\newcommand{\iconDimThreeD}{\raisebox{-0.5ex}{\includegraphics[width=1.5em]{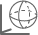}}}
\newcommand{\iconScopeInterior}{\raisebox{-0.5ex}{\includegraphics[width=1.2em]{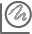}}}
\newcommand{\iconScopeExterior}{\raisebox{-1.0ex}{\includegraphics[width=1.8em]{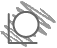}}}
\newcommand{\iconScopeSeparate}{\raisebox{-1.0ex}{\includegraphics[width=2em]{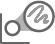}}}
\newcommand{\iconImgTwoD}{\raisebox{-0.5ex}{\includegraphics[width=1.3em]{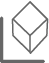}}}
\newcommand{\iconImgThreeD}{\raisebox{-0.5ex}{\includegraphics[width=1.7em]{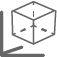}}}
\newcommand{\iconImgDecal}{\raisebox{-0.5ex}{\includegraphics[width=1.9em]{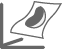}}}
\newcommand{\iconEncodingThematic}{\raisebox{-0.5ex}{\includegraphics[width=1.4em]{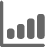}}}
\newcommand{\iconEncodingSpatial}{\raisebox{-0.5ex}{\includegraphics[width=1.7em]{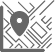}}}

\newcommand{\iconViewDependent}{\raisebox{-0.5ex}{\includegraphics[width=1.2em]{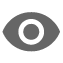}}}

\newcommand{\iconBoxPOS}{\raisebox{-0.5ex}{\includegraphics[width=1em]{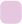}}}
\newcommand{\iconBoxShape}{\raisebox{-0.5ex}{\includegraphics[width=1em]{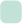}}}
\newcommand{\iconBoxDimensionality}{\raisebox{-0.5ex}{\includegraphics[width=1em]{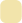}}}
\newcommand{\iconBoxScope}{\raisebox{-0.5ex}{\includegraphics[width=1em]{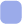}}}
\newcommand{\iconBoxImagery}{\raisebox{-0.5ex}{\includegraphics[width=1em]{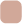}}}
\newcommand{\iconBoxEncoding}{\raisebox{-0.5ex}{\includegraphics[width=1em]{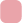}}}
\newcommand{\iconBoxView}{\raisebox{-0.5ex}{\includegraphics[width=1em]{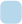}}}
\newcommand{\iconAsterisk}{\raisebox{-0.5ex}
{\includegraphics[width=1em]{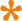}}}

\onlineid{0}

\vgtccategory{Research}
\vgtcpapertype{please specify}

\title{Spatially-Embedded Lens Visualization: A Design Space}

\author{Roberta Mota, Ehud Sharlin, Usman Alim}
\authorfooter{
\item
 Roberta Mota, Ehud Sharlin, and Usman Alim are with University of Calgary. E-mails: \{roberta.cabralmota, ehud, ualim\}@ucalgary.ca.
}

\abstract{ Lens visualization has been a prominent research area in the visualization community, fueled by the continuous need to mitigate visual clutter and occlusion resulting from the increase in data volume. 
Interactive lenses for \textit{spatial} data, particularly, challenge designers to conceive design strategies to support the analysis of high-density, multifaceted data with spatial referents.
Despite their relevance, there is a lack of systematic understanding regarding the various design elements that compose \textit{spatially-embedded} lens visualizations.
To fill in this gap, we unify these components under a common hood in the form of a design space, which we propose in this paper.
Building our knowledge on top of the initial insights gained from Tominski \textit{et al.}'s survey \cite{tominski2017}, we construct a design space spanning 7 dimensions through our analysis of 45 papers published in the visualization community over the past 15 years.
We describe each design dimension through representative examples and examine the range of design choices available within each, discussing their benefits and pitfalls that affect lens performance and usability.
In doing so, we offer a cohesive catalog of considerations for designers---both when examining existing lenses and when conceptualizing novel spatially-embedded lens visualizations.
We conclude by shedding light on regions of the design space that remain largely understudied, revealing open opportunities for future research.

} 


\CCScatlist{ 
 \CCScat{K.6.1}{Management of Computing and Information Systems}%
{Project and People Management}{Life Cycle};
 \CCScat{K.7.m}{The Computing Profession}{Miscellaneous}{Ethics}
}

\teaser{
  \centering
  \includegraphics[width=\linewidth]{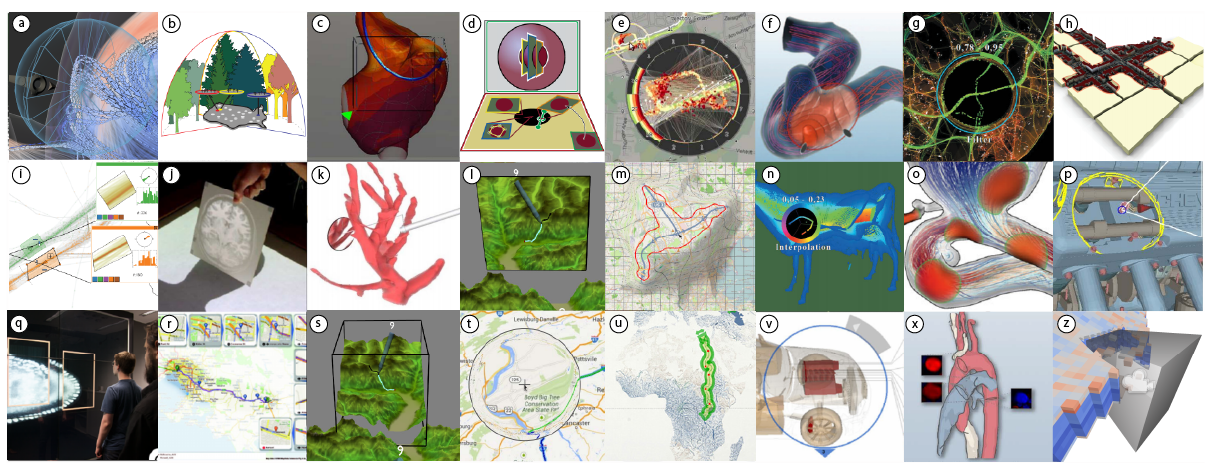}
  \caption{A collage of lens visualizations categorized across seven design dimensions.
  \texttt{\textcolor{black}{\textbf{Position, orientation \& scale}}}: interactive ($a$, $j$) and semi-automated ($t$).
  \texttt{\textcolor{black}{\textbf{Shape}}}: fixed ($k$, $h$), dynamic user-driven ($u$, $g$) and data-driven ($m$).
  \texttt{\textcolor{black}{\textbf{Dimensionality}}}: 2D ($q$, $v$), 2.5D ($b$, $d$), and 3D ($h$, $z$).
\texttt{\textcolor{black}{\textbf{Effect scope}}}: separate view ($r$, $x$), lens interior ($g$, $t$) and exterior ($a$, $i$).
  \texttt{\textcolor{black}{\textbf{Effect imagery}}}: 2D ($l$, $f$), 3D ($c$, $s$), and Decal ($o$).
  \texttt{\textcolor{black}{\textbf{Effect encoding}}}: thematic ($e$, $i$) and spatially-based ($q$, $m$).
  \texttt{\textcolor{black}{\textbf{Viewpoint Dependency}}}: independent ($p$, $a$) and view-dependent ($q$, $z$).}
  \label{fig:teaser}
}

\vgtcinsertpkg

\begin{document}


\firstsection{\textcolor{black}{Introduction}}
\maketitle

As data size is constantly increasing, visualization techniques have to deal with the problem of visual clutter and occlusion---\textit{occlusion} hides important information from the viewer, whereas \textit{clutter} shows too much visual information and makes the visualization more complex for the viewer to understand and gain insight from it.
The high data density and dimensionality associated with spatial visualizations, particularly, pose significant challenges in encoding all relevant information into a single visual imagery. 
To overcome the challenges of clutter and occlusion, the visualization community has embraced the idea of \textit{interactive lenses}---a family of focus+context techniques that provide on demand an alternative visual representation of a focus area of the screen, while the remaining regions convey context \textcolor{black}{\cite{tominski2017}}.

In the last decades, numerous lenses have been proposed in the visualization research, with a growing number in the field of immersive analytics which leverages immersive technologies for visual data analysis \textcolor{black}{\cite{tominski2017, CHEN2017}}.
Despite their relevance, there is still few work that provides a more systematic knowledge on the question of \textit{what are design components and choices to be considered when constructing visualization lenses?} This is particularly true in the case of lenses that address the challenges encountered in \textit{spatial} visualizations.

\vspace{3px}
\noindent{\textbf{Contribution}} To fill in this gap, we contribute in this paper a  \textbf{{design space}} of interactive lenses for spatial data, namely \textit{spatially-embedded lens visualizations}. Through our survey and analysis of 45 exemplary lenses published over the past 15 years in the visualization literature, we systematically formulate 7 design dimensions that capture the salient aspects of interactive lenses operating within spatial environments---contemplating both conventional and more immersive spaces.

\vspace{3px}

\noindent{\textbf{Outline}} In Section \ref{sec:DefinitionConceptLens}, we introduce the core concepts of lenses. In Section \ref{sec:designSpace}, we proceed to describe our design space by explaining each design dimension with examples, and outlining the available design choices within each dimension along with their trade-offs that influence lens performance or usability. 
In Section \ref{sec:caseStudies}, \textcolor{black}{we demonstrate the applicability and generative capacity of our conceptual schema through a detailed design use case.}  Lastly, in Section \ref{sec:discussionLimitations}, we leverage our design space to identify and discuss as-of-yet understudied design areas that reflect opportunities for future research. 

\section{\textcolor{black}{Lens Conceptual Model \& Definition}}
\label{sec:DefinitionConceptLens}
\vspace{-2px}

\begin{figure}[t!]
    \centering
    \includegraphics[width=1\linewidth]{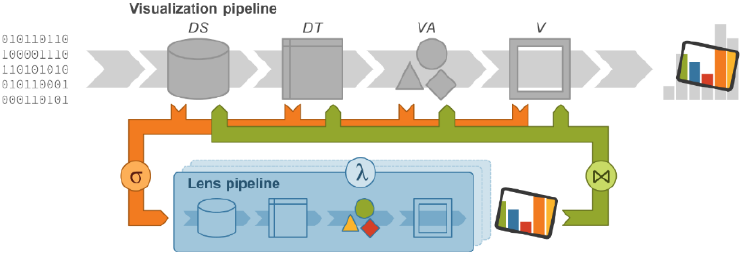}
    \vspace{-0.65cm}
    \caption{Conceptual model of visualization lens \cite{tominski2017}. The standard visualization pipeline describes how data are transformed from a data source (DS) via data tables (DT) and visual abstractions (VA) to a visualization view (V). Tominski \textit{et al.} models a lens as a pipeline embedded in the standard pipeline, and with three main stages: lens selection ($\sigma$), lens function ($\lambda$), and join ($\bowtie$).} 
    \label{fig:pipelineLenses}
    \vspace{-0.45cm}
\end{figure}

\noindent \textbf{Conceptual model} Tominski \textit{et al.} define a conceptual model of visualization lenses built upon the well-established visualization pipeline \textcolor{black}{\cite{tominski2017}}, as illustrated in \textcolor{black}{Fig. \ref{fig:pipelineLenses}}. Taking this as a basis, the lens pipeline has three main stages. First, the \textit{lens selection} ($\sigma$) captures what part of the data is to be affected by the lens. Second, the \textit{lens function} ($\lambda$) is the operator that defines how the lens modifies the visualization, and may depend on parameters that control the lens effect---\textit{e.g.} a zoom factor of a magnification lens, or an alpha or threshold value of a filtering lens. Lastly, the join ($\bowtie$) stage joins the \textit{lens effect} with the base visualization. We incorporate this conceptual model along with its terminology throughout our manuscript.

\vspace{5px}

\begin{figure}[t!]
    \centering
    \includegraphics[width=1\linewidth]{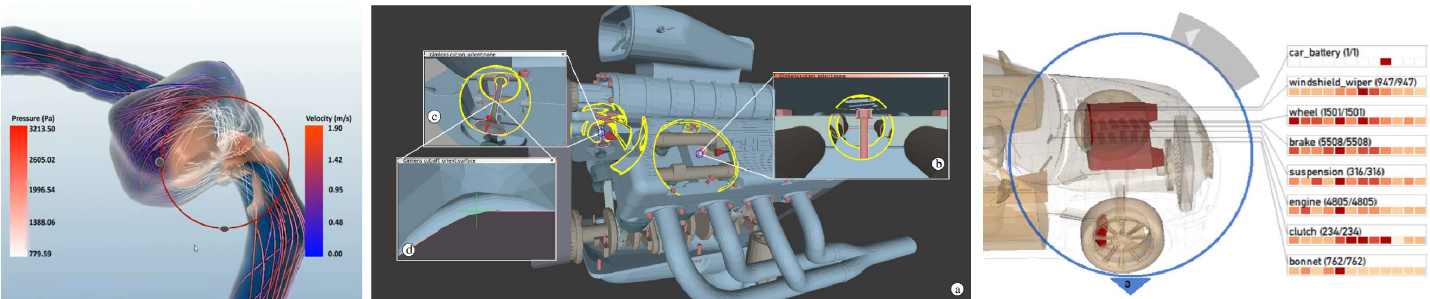}
    \vspace{-0.65cm}
    \caption{Examples of visualization lenses. A lens may \textit{alter} a base visualization to exhibit a different attribute or rendering style--\textit{e.g.} isosurface and streamline \cite{gasteiger2011} (\textit{left}). Also, a lens may not only \textit{suppress} occluding outer layers to reveal a focus object located inside a 3D model, but also \textit{enrich} with complementary views of the data in focus--\textit{e.g.} views with different magnification levels and varying viewpoints \cite{pindat2013} (\textit{middle}), or semantic views \cite{chang2013} (\textit{right}).} 
    \label{fig:exampleLenses}
    \vspace{-0.6cm}
\end{figure}

\noindent The definition of visualization lenses originates from the idea of a physical lens moved above an area of interest to temporarily magnify the visibility of details. However, this traditional interpretation has been significantly broadened in the context of visualization \textcolor{black}{\cite{tominski2017}}.
Beyond solely magnifying visual elements, lenses have been extended to include a range of functions for manipulating the visualization.
In general, a lens function can modify a base visualization by \textit{altering} existing content, \textit{suppressing} irrelevant content, \textit{augmenting} with new content, or even a combination thereof---see \textcolor{black}{Fig. \ref{fig:exampleLenses}}.

Furthermore, although lens effects are often transient by \textit{temporarily} manipulating the view of focus
areas, they may also be used to \textit{permanently} modify the visualization---\textit{e.g.} copy a visual element located under the lens boundary and paste copies of it in other parts of the visualization \textcolor{black}{\cite{bier1993, spindler2012}}.
Finally, lens functions are \textit{interactively parametrizable}. This means that users may interact with the lens to adjust parameters that control the lens effect while carrying out visual data exploration and analysis---\textit{e.g}. the zoom factor or even the mathematical function specifying the magnification distortion.

\vspace{5px}

\noindent \textbf{Definition} Considering the previously discussed aspects, we consider the following definition of visualization lenses: \textit{an interactively parametrizable spatial selection according to which a base visualization is modified to provide an alternative visual representation of the data in focus \footnote{Similar to the definition of Tominski \textit{et al}. \cite{tominski2017}}.}

\vspace{-3px}
\section{\textcolor{black}{Design Space}}
\label{sec:designSpace}
\vspace{-2px}

Although numerous visualization lenses have been introduced over the past decades, the question of \textit{what are design components and choices to be considered when constructing visualization lenses?} remains understudied, as few work have attempted to establish a structured space that documents lens techniques by describing them in well-defined design elements.
Establishing a design space is valuable since it allows us not only to systematically examine existing lenses but also to reason about new lens-based techniques. This section attempts to define such a design space of spatially-embedded visualization lenses.

An early taxonomy on lenses was proposed by Bier \textit{et al}. \cite{bier1994taxonomy}, and later extended by Tominsky \textit{et al}. \cite{tominski2017}. The latter taxonomy encompasses both \textit{practical} aspects that relate to lens usage, such as data types, user tasks, and display mediums, as well as \textit{conceptual} aspects primarily concerning the visual design of lenses. Here, we aim to extend the existing conceptual schema as listed in Table \ref{tab:tableTaxonomy}. To this end, we discuss the \texttt{\textcolor{black}{\textbf{Position, Orientation \& Scale}}} and \texttt{\textcolor{black}{\textbf{Shape}}} dimensions, which have been previously defined as properties driving lens selection \cite{tominski2017}.
The \texttt{\textbf{\textit{Effect Scope}}} dimension is defined as the area within the visualization where the lens effect is applied, and is a refinement of the \textit{ Effect Extent} class from the previous conceptual schema.
The remaining dimensions---namely, \texttt{\textcolor{black}{\textbf{Dimensionality}}}, \texttt{\textcolor{black}{\textbf{Effect Imagery}}}, \texttt{\textcolor{black}{\textbf{Effect Encoding}}}, and \texttt{\textcolor{black}{\textbf{Viewpoint Dependency}}}, have not been formally defined in the literature so far.
We explain each of the $7$ aforementioned design dimensions with examples, and outline the set of possible design choices within each dimension along with their trade-offs that influence lens performance or usability.

\begin{table}[b!]
  \begin{center}
  \vspace{-0.5cm}
  \centering
    \begin{tabular}{ m{0.35\linewidth} m{0.65\linewidth} }
       
      \textit{{Effect Class}}: &    Suppress, alter, enrich \\
      \vspace{0.1cm} 
      \textit{{Effect Extent}}: &   Lens interior, side effects, separate view \\
      \vspace{0.1cm}
      \textit{{Adjustability}}: &   None, interactive, self-adjusting \\
      \vspace{0.1cm}
      \textit{{Selection Stage ($\sigma$)}}: &  \textit{DS}, \textit{DT} , \textit{VA}, \textit{V} \\
      \vspace{0.1cm}
      \textit{{Join Stage ($\bowtie$)}}: & \textit{DS}, \textit{DT} , \textit{VA}, \textit{V} \\      

      \hline & \\[-1.5ex]

      \textit{Pos, Ori \& Scale}: & Interactive, semi-automated \\
      \vspace{0.1cm}
      \textit{Shape}:                & Fixed, user- or data-driven dynamic \\
      \vspace{0.1cm}
      \textit{Dimensionality}:       & 2D, 2.5D, 3D \\
      \vspace{0.1cm}
      \textit{Effect Scope}:         & Interior, exterior, separate view \\
      \vspace{0.1cm}
      \textit{Effect Imagery}:       & 2D, 3D, Decal \\
      \vspace{0.1cm}
      \textit{Effect Encoding}:      & Thematic, spatially-based \\
      \vspace{0.1cm}
      \textit{Viewpoint Dependency}:    & Invariant, view-dependent \\
    
    \end{tabular}
      \vspace{-0.05cm}
      \caption{ Our design space comprises seven dimensions (\textit{bottom}) and expands upon Tominski \textit{et al.} 's five-axis conceptual schema \cite{tominski2017} (\textit{top}).
  }
    \vspace{0.25cm}
    \vspace{-1.5cm}
    \label{tab:tableTaxonomy}
  \end{center}
\end{table}

\vspace{5px}
\noindent{\textbf{Sampling criteria}} To collect relevant papers for our review we established three inclusion criteria: 

\vspace{4px}
\noindent $C_1$. Papers published in the past $15$ years, from $2006$ to $2020$.

\vspace{1px}
\noindent $C_2$. Papers presenting techniques that fit into our definition of visualization lens---see \textcolor{black}{Section \ref{sec:DefinitionConceptLens}}, which included techniques that call themselves lenses or have lens-like characteristics.

\vspace{1px}
\noindent $C_3$. Papers demonstrating lens techniques that address spatial data in either 2D or 3D spaces---by \textit{spatial data} we mean data with inherent width, height, or depth, where the relative positioning, length, shape, \textit{etc.} are meaningful for users.
Although our discussion in this paper focuses on 3D data, our decision to consider both spatial 2D and 3D visualizations was based on their shared need to account for the spatial nature of the data.
Consequently, they share concerns and constraints when designing lenses---\textit{e.g.} accurate spatial perception, spatially coherent transitions between focus and context areas when applying lens effects, as well as occlusion and clutter management when representing multidimensional data with specific spatial referents.
In doing so, we expanded our paper sample to facilitate extracting common lens design strategies and formulate our design space.
 
\vspace{1px}
\noindent $C_4$. Papers proposing lens-based visualizations in either conventional or more immersive environments.
Our decision to include immersive lenses was aimed at forming a more comprehensive taxonomy that incorporates design aspects accounting for emerging displays, thereby contributing to the growing field of immersive analytics.

\begin{figure*}[t!]
  \centering
    \includegraphics[width=1\textwidth]{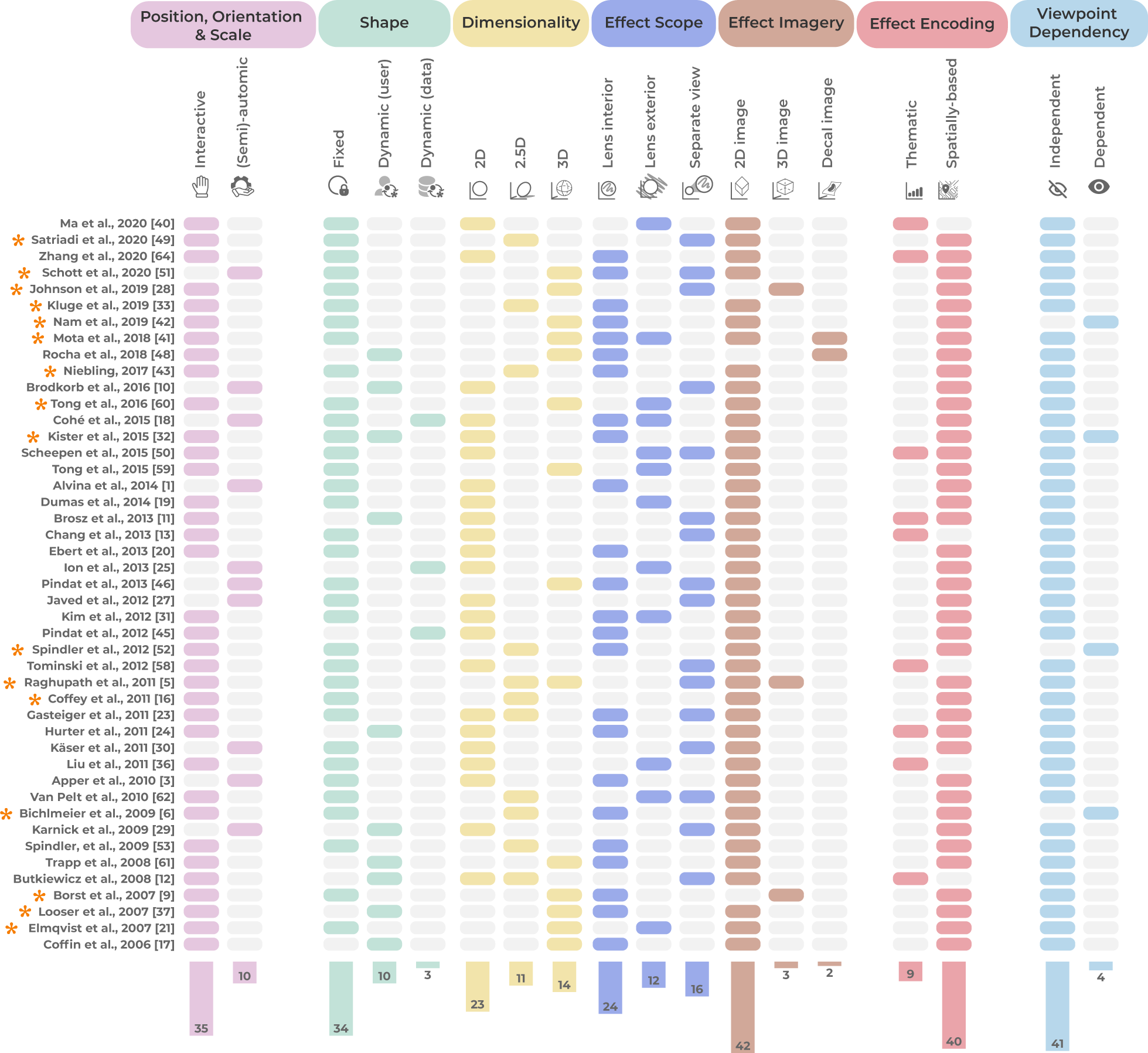}
   \vspace{-0.65cm}
 \caption{A summary of the coded papers that formed the \textcolor{black}{$45$} surveyed papers from $2006$ to $2020$, serving as a basis for our design space. The cells with different colors represent \texttt{\textcolor{black}{\textbf{Position, Orientation, and Scale}}} (\iconBoxPOS), \texttt{\textcolor{black}{\textbf{Effect Scope}}} (\iconBoxScope), \texttt{\textcolor{black}{\textbf{Shape}}} (\iconBoxShape), \texttt{\textcolor{black}{\textbf{Dimensionality}}} (\iconBoxDimensionality), \texttt{\textcolor{black}{\textbf{Effect Imagery}}} (\iconBoxImagery), \texttt{\textcolor{black}{\textbf{Effect Encoding}}} (\iconBoxEncoding), and \texttt{\textcolor{black}{\textbf{Viewpoint Dependency}}} (\iconBoxView) dimensions. Some references contain more than one colored cell within a single dimension, given its non-mutually exclusive design choices. Papers presenting lens techniques within immersive spaces are marked with an asterisk (\iconAsterisk).
 }
 \label{fig:codingDesignSpace}
  \vspace{-.6cm}
\end{figure*}

\vspace{5px}
\noindent{\textbf{Methodology}} To find relevant papers for our review we used a snowball sampling technique starting with the seminal paper by Bier \textit{et al.} \cite{bier1993}, which first introduced the concept of interactive lenses for visualization.
From this starting point, we recursively scanned references for further papers about lenses, and we stopped searching and collecting papers once this process reached saturation and no new relevant papers were identified.
From the collected papers, we derived our design space through open coding, based on consensus among all authors.
The resulting design codes and coded papers can be seen in \textcolor{black}{Fig. \ref{fig:codingDesignSpace}}.

\vspace{1px}
\noindent Drawing from our literature sampling and coding, we analyzed  \textcolor{black}{$45$} papers and derived a design space with $7$ design dimensions, systematically describing many design components and choices to be considered when constructing visualization lenses for spatial data:
\texttt{\textcolor{black}{\textbf{Position, Orientation \& Scale}}}, \texttt{\textcolor{black}{\textbf{Shape}}}, \texttt{\textcolor{black}{\textbf{Dimensionality}}}, \texttt{\textcolor{black}{\textbf{Effect Scope}}}, \texttt{\textcolor{black}{\textbf{Effect Encoding}}}, \texttt{\textcolor{black}{\textbf{Effect Imagery}}}, and \texttt{\textcolor{black}{\textbf{Viewpoint Dependency}}}.
Although not exhaustive, our design space forms a systematic understanding of available design components, choices, and their trade-offs---summarized in Table \ref{tab:summaryTradeOffs}---in order to guide the development  of new lens visualizations.
The following subsections catalog our design space by explaining each design dimension along with relevant examples.

\vspace{-3px}
\subsection{\textcolor{black}{Position, Orientation \& Scale}}
\vspace{-2px}

Because of the tight interrelation between position, orientation, and scale, we consider them together within a single design dimension. These lens parameters primarily specify the lens selection.

Most lens techniques supply users \texttt{\textcolor{colorPOS}{\textbf{full interactivity}}} \iconPOSInteractive\ for setting the lens position, orientation, and scale. 
Tominski \textit{et. al} proposed a circular lens where users interactively set its position and radius, delimiting the lens selection over spatiotemporal data on 2D maps \cite{tominski2012}. 
Lens orientation, along with position and scale, can influence focus selection.
Dumas \textit{et al.} proposed an angular-based lens technique for selecting curves within 2D visualizations---\textit{e.g.}  spatiotemporal movement data on maps. The lens' circular selection is specified not only by its position and diameter but also by direction and angular tolerance \cite{dumas2014vectorlens}.
Orientation is often overlooked in 2D spaces but becomes more essential when working with 3D data. 
By adjusting orientation, the lens can align its geometry with the underlying 3D visualization---\textit{e.g}. aligning a spatially-tracked mirroring lens with intertwined blood vessels  \textcolor{black}{\cite{bichlmeier2009}}.

Although most surveyed lenses are fully interactive, some rely on  \texttt{\textcolor{colorPOS}{\textbf{(semi-)automated}}} \iconPOSAutomatic\ control based on the underlying data \textcolor{black}{\cite{cohe2015, alvina2014, appert2010}}.
\textcolor{black}{Semi-automation may be useful particularly in scenarios where domain knowledge is encoded into the automation logic, guiding users towards relevant data features---\textit{e.g.} flow patterns.
It can also enhance precision in spatially-constrained visualizations or in tasks requiring more fine-grained lens manipulations, like route-following.}
To illustrate, Alvina \textit{et al}. introduced a magnification lens to reveal details while users steer along map routes \textcolor{black}{\cite{alvina2014}}. 
Although it is ideal to keep the lens center close to the route---since fisheye distortions aggravates with distance from the center, traditional magnification lenses often fail and let paths “slip off” the lens.
To address this, the lens automatically adjusts its position its position based on the route geometry, making routes easier to follow.
\textcolor{black}{Although \texttt{\textcolor{colorPOS}{\textbf{semi-automated}}} lenses can reduce user effort and enhance precision, they may reduce user agency if the underlying algorithms misalign with users' intent. Additionally, they may highlight irrelevant features if heuristics fail to generalize across datasets.}

\begin{figure}[b!]
    \vspace{-0.4cm}
    \centering
    \includegraphics[width=1\linewidth]{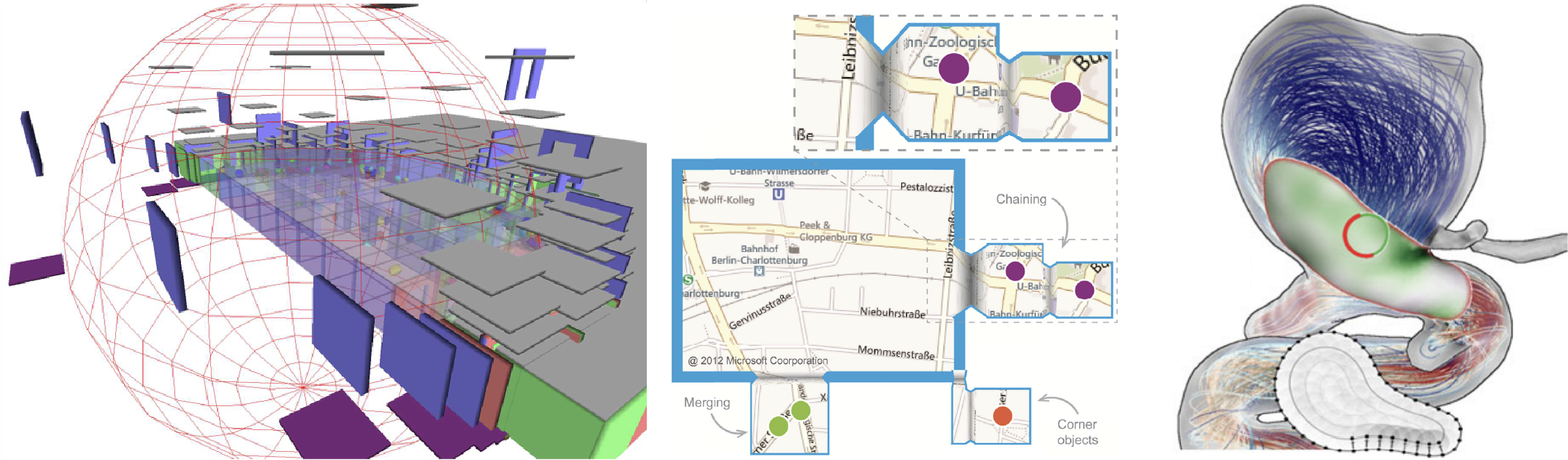}
    \vspace{-0.7cm}
    \caption{From \textit{left} to \textit{right}: a \texttt{\textcolor{colorShape}{\textbf{fixed-shape}}} distortion lens used to separate structures within an architectural model \cite{elmqvist2007occlusion}; a \texttt{\textcolor{colorShape}{\textbf{data-driven}}} lens shape dynamically adapts to retain off-view moving cars in focus using a paper folding metaphor \cite{ion2013}; lassoing operations define \texttt{\textcolor{colorShape}{\textbf{user-driven dynamic shapes}}} that display hemodynamic surface attributes \cite{rocha2018}. } 
    \label{fig:exampleShape}
\end{figure}

\vspace{-3px}
\subsection{\textcolor{black}{Shape}}
\vspace{-2px}

The shape of a lens refers to the geometry used for the lens selection, and can be classified as either \texttt{\textcolor{colorShape}{\textbf{fixed}}} or \texttt{\textcolor{colorShape}{\textbf{dynamic}}}---see Fig. \ref{fig:exampleShape}. 

Most lenses proposed in the literature are of \texttt{\textcolor{colorShape}{\textbf{fixed}}} \iconShapeFixed\ shape.
While they typically use simple, primitive forms---\textit{e.g}. circular, rectangular, box \textcolor{black}{\cite{nam2019, johnson2019, gasteiger2011}}, there are no restrictions preventing rigid, non-standard shapes. For instance, Trapp \textit{et al}. proposed a lens to display different levels of structural abstraction in 3D city models \textcolor{black}{\cite{trapp2008}}. The lens technique uses predefined 3D models as custom lens shapes, created using external modeling software.
\textcolor{black}{Fixed lens shapes typically offer simplicity and familiarity. They are easy to implement and interact with, which can lower the learning curve. However, they may not adequately conform to irregular data structures, resulting in poor coverage of relevant features, inclusion of irrelevant ones, and mental burden as users themselves must disambiguate them. }

Some lens techniques adapt their shape \texttt{\textcolor{colorShape}{\textbf{dynamically}}}, either \texttt{\textcolor{colorShape}{\textbf{by the user}}} \iconShapeDinamicUser\ or \texttt{\textcolor{colorShape}{\textbf{by the underlying data}}} \iconShapeDinamicData. 
In the former case, lens shapes are typically set by users through sketching, brushing, or lassoing (Fig. \ref{fig:exampleShape}--\textit{right}) \textcolor{black}{\cite{rocha2018, brosz2013, hurter2011}}.
Brosz \textit{et al}. proposed a lens-like technique where shapes can be sketched over 2D maps \textcolor{black}{\cite{brosz2013}}. Users draw an \textit{origin} shape that matches a target feature, such as a route, and a \textit{destination} shape. The visual content within the origin is then morphed to fit the destination form. 
Kister \textit{et al}. proposed a body-centric lens whose shape matches the body silhouette, like a shadow cast by the body onto the screen \cite{kister2015}.
\textcolor{black}{\texttt{\textcolor{colorShape}{\textbf{User-driven dynamic}}} shapes provide greater versatility and precision for selecting irregular features, better reflecting user intent. The trade-off is increased interaction effort---\textit{e.g.} manual drawing in dense 3D spaces can become cumbersome.}
For the latter case, \texttt{\textcolor{colorShape}{\textbf{data-driven dynamic}}} lenses automatically adapt their shape to match data features---\textit{e.g.} boundaries, clusters, paths. 
Pindat \textit{et al}. proposed a 2D magnification lens that implicitly adapts to the geometry of a focus object underlying the mouse cursor---\textit{e.g}. a topographic path or area. 
This shape morphing aims to provide less-distorted magnified images compared to traditional radial fisheye lenses \textcolor{black}{\cite{pindat2012}}.
Ion \textit{et al}. introduced magnification lenses with dynamic shapes to maintain focus of moving data targets on a map that become off-view of the lens boundary (Fig. \ref{fig:exampleShape}--\textit{middle}) \cite{ion2013}.
\textcolor{black}{Data-driven automation can reduce user effort, enable automated feature discovery, and ensure lenses align meaningfully with the data---thus improving interpretability. However, these lenses may behave unpredictably when the data is noisy or lacks clear structure, making it difficult for users to understand or control the logic behind the autogenerated shape configuration. Therefore, implementing such behavior requires careful data preprocessing and logic calibration or transparency to ensure reliability.}

\begin{figure}[t!]
    \centering
    \includegraphics[width=1\linewidth]{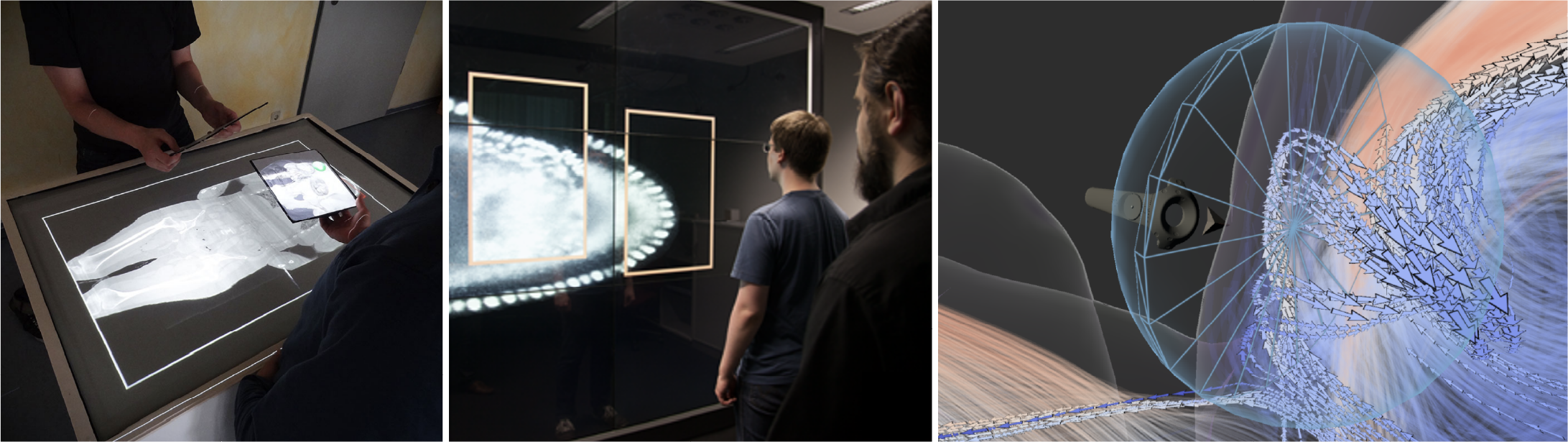}
    \vspace{-0.7cm}
    \caption{ From \textit{left} to \textit{right}: a \texttt{\textcolor{colorDimensionality}{\textbf{2.5D}}} lens depicting photo-realistic details of a 3D human anatomic model \cite{spindler2012}; \texttt{\textcolor{colorDimensionality}{\textbf{2D}}} lenses showcasing different magnification levels of a 3D dental arcade model \cite{kister2015}; a \texttt{\textcolor{colorDimensionality}{\textbf{3D}}} lens highlights hemodynamic flow passing through its region \cite{mota2018}. } 
    \label{fig:exampleDimensionality}
    \vspace{-0.6cm}
\end{figure}


\vspace{-2px}
\subsection{\textcolor{black}{Dimensionality}}
\vspace{-2px}

The dimensionality is classified either as \texttt{\textcolor{colorDimensionality}{\textbf{2D}}}, \texttt{\textcolor{colorDimensionality}{\textbf{2.5D}}}, or \texttt{\textcolor{colorDimensionality}{\textbf{3D}}}---see Fig.~\ref{fig:exampleDimensionality}, and relates to the other geometric parameters of lenses that drive the lens selection---\textit{i.e.} position, orientation, scale, and shape. 

\texttt{\textcolor{colorDimensionality}{\textbf{\textit{2D lenses}}}} \iconDimTwoD\ are the most commonly used and have polygonal shapes—-\textit{e.g.} circular or square, which are placed and manipulated in screen space.
For instance, Tim and Elmqvist proposed the use of physical lenses made of thin, transparent paper sheets on top of tabletop surfaces \cite{kim2012embodied}. Their physicality affords them to be intuitively overlapped, creating composite lens effects on the underlying visualization.
Although \texttt{\textcolor{colorDimensionality}{\textbf{2D}}} lenses have been used in 3D visualizations \cite{ma2020, kister2015, hurter2011, chang2013}, \textcolor{black}{their 2D nature constrains the lens manipulations to the view plane and leads to two primary limitations when handling 3D data: an inability to carry out a 3D selection, and an absence of spatial correlation between the 2D alignment of the lens and the underlying 3D visualization.}

\texttt{\textcolor{colorDimensionality}{\textbf{2.5D lenses}}} \iconDimTwoHalfD\ consist of \texttt{\textcolor{colorDimensionality}{\textbf{2D}}} lenses embedded in 3D spaces, thereby augmenting lens manipulations to 3D \cite{nam2019, kluge2019virtual, gasteiger2011, bichlmeier2009}. \textcolor{black}{Although this alleviates the issue of spatial correlation, it increases interaction effort to place and align the lenses according to the underlying data. In addition, the flat nature of \texttt{\textcolor{colorDimensionality}{\textbf{2.5D}}} lenses constraint the spatial selection to only a slice of the 3D data.} To illustrate, Spindler \textit{et al.} introduced a spatially aware, handheld paper sheet lens to improve lens manipulations in 3D spaces above tabletop displays \cite{spindler2009}---\textit{e.g.} users could arbitrarily slice an MRT scan of a human torso in different poses.

\texttt{\textcolor{colorDimensionality}{\textbf{3D lenses}}} \iconDimThreeD\ consist of 3D volumes---\textit{e.g.} a sphere,
which enable volumetric spatial selections of the 3D data \cite{johnson2019, trapp2008, borst2007volume}. \textcolor{black}{Due to their three-dimensional nature, the lens manipulations are prone to suffer from interaction effort similar to \texttt{\textcolor{colorDimensionality}{\textbf{2.5D}}} lenses. Additionally, such lenses may not closely align with surface geometries due to their typically \texttt{\textcolor{black}{\textbf{fixed}}} shapes; this can lead to perceptual issues such as the lens itself occluding parts of visualization}---especially those of intricate geometry. Viega \textit{et al.} first introduced \texttt{\textcolor{colorDimensionality}{\textbf{3D}}} lenses to limit the lens effect to a finite subvolume of interest \cite{viega19963}. The lens effect is built from the intersection of GPU clipping planes for revealing hidden structures of 3D datasets.

\begin{figure*}[t!]
    \centering
    \includegraphics[width=1\linewidth]{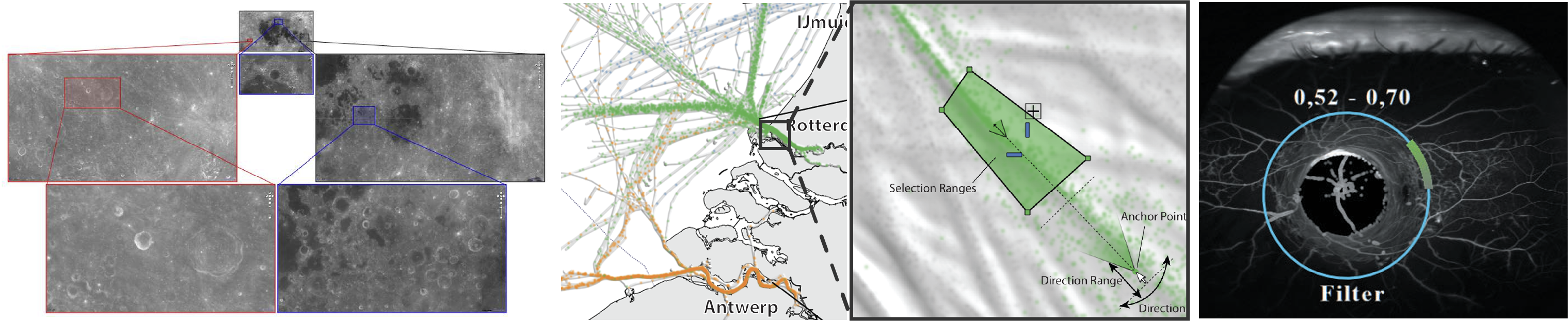}
    \vspace{-.75cm}
    \caption{From \textit{left} to \textit{right}: Color-coded and connecting lines being used to highlight parent-child relationships among multiple \texttt{\textcolor{black}{\textbf{rectangular}}}, magnified  lens images depicted in \texttt{\textcolor{colorScope}{\textbf{separate views}}} \cite{javed2012}; In a density map, a lens selects traffic flows based on their position and orientation, highlighting them as colored particles moving along the trajectories that go \texttt{\textcolor{colorScope}{\textbf{beyond the lens boundary}}} \cite{scheepens2015}. ; A \texttt{\textcolor{black}{\textbf{circular}}} lens applied to a grayscale angiography image, in which low-brightness pixels located \texttt{\textcolor{colorScope}{\textbf{inside the lens}}} are gradually displaced toward the lens border \cite{hurter2011}.} 
    \label{fig:exampleScope}
    \vspace{-.6cm}
\end{figure*}

\begin{figure}[b!]
    \centering
    \vspace{-.55cm}
    \includegraphics[width=1\linewidth]{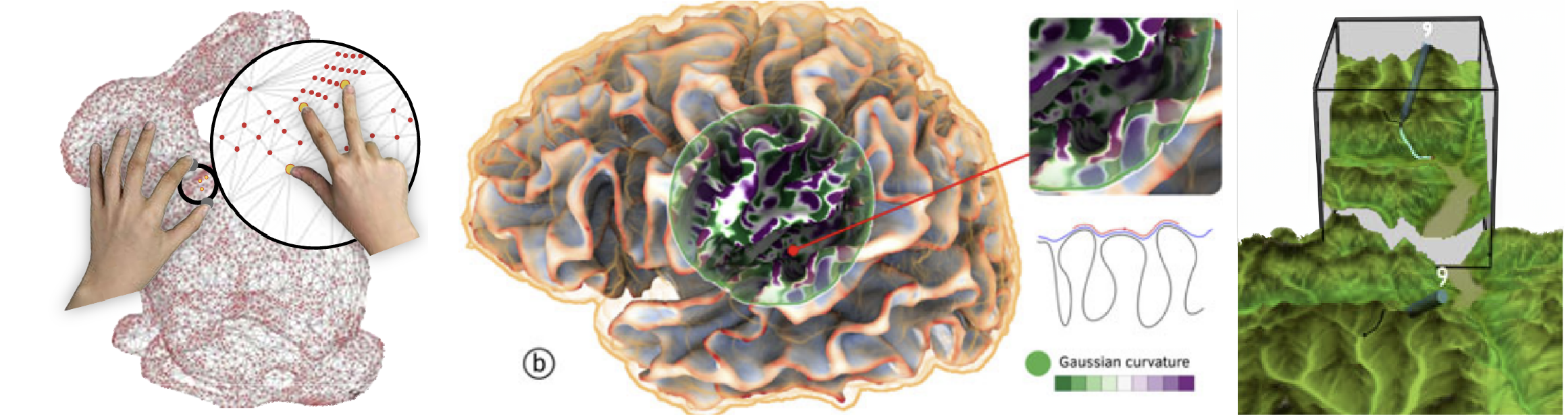}
    \vspace{-.75cm}
    \caption{ From \textit{left} to \textit{right}: A \texttt{\textcolor{black}{\textbf{circular}}} lens generates a \texttt{\textcolor{colorImagery}{\textbf{2D image}}} that enlarges the vertices of a 3D bunny model \cite{kaser2011fingerglass}; A \texttt{\textcolor{colorImagery}{\textbf{decal}}}-based lens depicts color-coded attributes on surfaces of intricate geometry such as the human brain \cite{rocha2018}; A \texttt{\textcolor{colorImagery}{\textbf{3D imagery}}} being used to reveal hidden surface regions as users trace paths on a geological model \cite{bezawada2011investigation}. }
    \label{fig:exampleImagery}
    
\end{figure}

\vspace{-2px}
\subsection{\textcolor{black}{Effect Scope}}
 \vspace{-2px}
 
The lens effect scope refers to the area within the visualization where the lens applies its effect, which can be classified as \texttt{\textcolor{colorScope}{\textbf{lens interior}}}, \texttt{\textcolor{colorScope}{\textbf{lens exterior}}}, or \texttt{\textcolor{colorScope}{\textbf{separate view}}}---see Fig. \ref{fig:exampleScope}.

Traditionally, a lens function only applies to visual elements confined inside its shape bounds—-that is, the lens effect scope is the \texttt{\textcolor{colorScope}{\textbf{lens
interior}}} \iconScopeInterior\ \cite{nam2019, hurter2011, appert2010}.
\textcolor{black}{Restricting the lens effect to its interior maintains a clear distinction between focus and context. It also localizes visual changes, minimizing disruption in the surrounding visualization.} To illustrate, Hurter \textit{et al.} presented a lens that selects a specific spatial and attribute-related data range. The lens preserves the data in focus but continuously deforms in-lens data outside the selection range, moving it toward the lens border (Fig. \ref{fig:exampleScope}--\textit{right}) \cite{hurter2011}.
\textcolor{black}{Although interior effects offer localized focus+context visualization, it may obscure broader patterns or connections extending beyond the lens boundary---especially in tasks requiring spatial continuitiy, like tracing trajectories.}

A few lens effects impact visual elements located not only inside the lens boundary but also beyond it; therefore, the effect scope is considered to be the \texttt{\textcolor{colorScope}{\textbf{lens exterior}}} \iconScopeExterior\ \cite{mota2018, scheepens2015, dumas2014vectorlens}. In such cases, the lens boundary is primarily used to define a selection based on a local behavior of interest in the underlying data---\textit{e.g.} to select trajectories based on a local directionality. 
To illustrate, Mota \textit{et al.} introduced a lens that highlights streamlines passing through the lens boundary \textit{and} having slopes locally oriented approximately in the same direction as the lens  (Fig. \ref{fig:exampleDimensionality}--\textit{right}) \textcolor{black}{\cite{mota2018}}. 
Tong \textit{et al.} proposed deformation-based lenses to clear the line of sight between the viewer and  3D glyphs in focus \cite{tong2016}. The lenses remove occluding glyphs situated within the lens boundary by displacing them aside in the surrounding context.
\textcolor{black}{\texttt{\textcolor{colorScope}{\textbf{Exterior effects}}} are also useful when users wish to explore patterns emerging across multiple foci---by composing two or more lenses to build cascaded selections or effects across spatial regions \cite{kruger2013trajectorylenses}.}
For instance, Scheepens \textit{et al.} proposed a lens for trajectories in 2D visualizations  (Fig. \ref{fig:exampleScope}--\textit{middle}) \textcolor{black}{\cite{scheepens2015}}. The technique supports composing multiple lenses to enable compound selections of more convoluted traffic flows of interest.
\textcolor{black}{\texttt{\textcolor{colorScope}{\textbf{Exterior-scope}}} lenses, however, can weaken spatial associations between the lens and its effect, creating ambiguity in attributing effects to their originating lenses.}

Some lenses render their effects in independent visual spaces by generating entirely new views separate from their source selection.
They commonly showcase either a duplicate of the visual content enriched with additional elements, or an alternative visual encoding that may not be suitable in the originating view.
This effect scope is referred to as \texttt{\textcolor{colorScope}{\textbf{separate view}}} \iconScopeSeparate\ \cite{pindat2013, brodkorb2016overview, scheepens2015, javed2012, tominski2012}.
\textcolor{black}{Lenses whose effects are displayed across separate views enable side-by-side comparisons; however, careful layout design is needed as to avoid becoming cognitive demanding by requiring users to mentally relate spatially-detached views.
Therefore, a key design consideration is how to spatially relate the \texttt{\textcolor{colorScope}{\textbf{separate views}}} to their corresponding lenses}---\textit{e.g.} via visual links or color coding \cite{scheepens2015}.
Karnick \textit{et al.} proposed a lens that uses a layout algorithm to determine optimal placement of focused views near referent locations along a selected geographical path in a 2D map \textcolor{black}{\cite{karnick2009route}}. The design priorities of the layout algorithm were to prevent the lens views from occluding the base map visualization and to maintain proximity to their associated points of interest.
\textcolor{black}{Furthermore, having separate views enables the definition of cascading effects, where each lens presents a distinct data transformation while preserving the history of prior lens stages.}
Javed \textit{et al}. proposed a lens-like technique that progressively constructs a hierarchical tree view from the user-defined, cascaded focus regions in a 2D map (Fig. \ref{fig:exampleScope}--\textit{left}) \cite{javed2012}.
Subsequent tree levels display views with higher magnification degrees, and the layout algorithm concentrates on forming clear parent-child relations to aid navigation along the cascaded hierarchy and on positioning views at a given tree level side by side to enable easier comparisons. 


\vspace{-5px}
\subsection{\textcolor{black}{Effect Imagery}}
\vspace{-2px}

The effect imagery refers to the visual image generated by the lens function, and can be classified as \texttt{\textcolor{colorImagery}{\textbf{2D}}} , \texttt{\textcolor{colorImagery}{\textbf{3D}}}, or \texttt{\textcolor{colorImagery}{\textbf{Decal images}}}---see Fig.~\ref{fig:exampleImagery}.

In \texttt{\textcolor{colorImagery}{\textbf{2D lens images}}} \iconImgTwoD\, the imagery generated as the lens effect consists of projected images, which are rendered either directly onto the view plane \cite{gasteiger2011, hurter2011} or onto a separate render texture to display the lens image as a \texttt{\textcolor{black}{\textbf{separate view}}} \cite{nam2019, pindat2013, bichlmeier2009, bezawada2011investigation}.
As an example, Pindat \textit{et al.} proposed a lens that renders its \texttt{\textcolor{colorImagery}{\textbf{projected}}} image in a \texttt{\textcolor{black}{\textbf{separate viewport}}}, which users can move and rescale. The \texttt{\textcolor{colorImagery}{\textbf{2D}}} lens image contains the same visual elements rendered within the lens boundary, but with a different magnification level and viewpoint \cite{pindat2013}.
\textcolor{black}{A limitation of \texttt{\textcolor{colorImagery}{\textbf{2D}}} images is providing only a monocular view, which may result in reduced spatial perception of depth and shape of the 3D data visualization. Therefore, an important design consideration is whether and how to use such images for visualization tasks that rely on shape and depth information---\textit{e.g.} identifying isolated blood vessel pathways may be challenging, as they could appear ambiguous and not clearly discernible from a static projected image due to overlaps, occlusion, and the lack of depth perception in a \texttt{\textcolor{colorImagery}{\textbf{2D}}} lens image \cite{bichlmeier2009}}. In this sense, a \textit{view-dependent} \texttt{\textcolor{colorImagery}{\textbf{2D}}} image may assist restoring the lost spatial perception, as motion parallax effects manifest when projections are continuously rendered in real time \cite{nam2019, bichlmeier2009}.

\texttt{\textcolor{colorImagery}{\textbf{3D images}}} \iconImgThreeD\ consist of volumetric visual images, possibly rendered as a separate volume and enriched with additional visual elements \cite{borst2007volume, johnson2019, bezawada2011investigation}---similar to a world-in-miniature \cite{stoakley1995virtual}.
As an example, Johnson \textit{et al}. presented a technique for exploring 4D cardiac data using a global-to-local approach. Each lens selection generates a magnified \texttt{\textcolor{colorImagery}{\textbf{3D}}} duplicate displayed in a \texttt{\textcolor{black}{\textbf{separate view}}}, and these views are automatically arranged in a juxtaposed grid. This enables visual comparison of key features across scales and points of view while minimizing irrelevant regions.
\textcolor{black}{Due to its volumetric nature, \texttt{\textcolor{colorImagery}{\textbf{3D}}} images preserve spatial perception better compared to \texttt{\textcolor{colorImagery}{\textbf{2D}}} projected images. However, rendering and interacting with \texttt{\textcolor{colorImagery}{\textbf{3D}}} imagery can be significantly more computationally costly than with \texttt{\textcolor{colorImagery}{\textbf{2D}}} imagery. 
This is particularly concerning  as immersive visualizations demand both high resolution and high frame rates, posing a greater challenge compared to visualizations on conventional displays, where lower frames rates and intermittent pauses for computations or data loading are more tolerable.}
To mitigate latency challenges, Johnson \textit{et al}. proposed a combination of data sampling strategies and clipped volume rendering to display volumetric images at interactive frame rates specifically for use in virtual reality \cite{johnson2019}. Borst \textit{et al}. proposed a data tiling technique to speed up rendering of multiple volumetric images of focus regions on topographic surfaces \cite{borst2007volume}.

\texttt{\textcolor{colorImagery}{\textbf{Decal images}}} \iconImgDecal\ conform to surface areas of interest. This conforming behavior \textcolor{black}{not only establishes a strong spatial correlation between the lens image and its referent location but also mitigates perceptual issues, such as the lens image itself occluding parts of the base visualization}---particularly those of intricate geometry.
To illustrate, Rocha \textit{et al.} proposed \texttt{\textcolor{colorImagery}{\textbf{decal}}} lenses for multivariate data visualization on surfaces \textcolor{black}{\cite{rocha2018}}. The lens image is built from the intersection of a sphere with a surface. The \texttt{\textcolor{black}{\textbf{sphere}}} itself does not define the area where the lens effect appears; rather, this area corresponds to the portion of the surface that lies within the sphere. This leads to a lens image that conforms to the surface geometry, similar to a decal.
Mota \textit{et al}. later proposed a hybrid lens that acts on different data types that commonly co-exist in 3D visualizations: streamlines and surfaces \cite{mota2018}. The lens integrates two categories of lense images---\texttt{\textcolor{colorImagery}{\textbf{3D}}} and \texttt{\textcolor{colorImagery}{\textbf{Decal}}}---to become a versatile lens that applies different effects according to the underlying data type in focus---non-surface and surface data, respectively.

\setlength{\tabcolsep}{2pt}  
\begin{table*}[b!]
\centering
\small
\vspace{-.4cm}

    \begin{tabular}{@{}p{0.15\linewidth}p{0.15\linewidth}p{0.15\linewidth}p{0.035\linewidth}p{0.15\linewidth}p{0.15\linewidth}p{0.01\linewidth}p{0.15\linewidth}@{}}
    \texttt{\textcolor{colorDimensionality}{\textbf{\large{Dimensionality}}}} &&&& \multicolumn{2}{l}{\texttt{\textcolor{colorEncoding}{\textbf{\large{Effect Encoding}}}}} & & \texttt{\textcolor{colorImagery}{\textbf{\large{Effect Imagery}}}} \\
    \cline{1-3}\cline{5-6} \cline{8-8} 
    \vspace{1em} & & \\[-1.5ex]

    \texttt{\textbf{2D}} & \texttt{\textbf{2.5D}} & \texttt{\textbf{3D}} & & \texttt{\textbf{Thematic}} & \texttt{\textbf{Spatial}} & &\texttt{\textbf{2D Image}}\\

    Standard 2D interaction* & Spatial interaction* & Volumetric selection* & & Higher-level reasoning* & Spatial reasoning* & & Familiar representation*\\
    Simpler lens layering* & Slice-only selection & Higher interaction effort & & Abstract-physical gap & Anchored visual cues* & & Lower rendering cost* \\
    Lacks spatial correlation &  & Misaligned, rigid selection  & & & Relies on user decoding & & Low spatial perception \\
     &  &   & & & Prone to clutter & &  \\
    \end{tabular}

\vspace{0.25em} 

    \begin{tabular}{@{}p{0.15\linewidth}p{0.15\linewidth}p{0.15\linewidth}p{0.035\linewidth}p{0.15\linewidth}p{0.15\linewidth}p{0.01\linewidth}p{0.15\linewidth}@{}}
    \texttt{\textcolor{colorShape}{\textbf{\large{Shape}}}} &  &  & &\multicolumn{2}{l}{\texttt{\textcolor{colorView}{\textbf{\large{Viewpoint Dependency}}}}} & & \\
    \cline{1-3}\cline{5-6}
    \vspace{1em} & & \\[-1.5ex]

     \texttt{\textbf{Fixed}} & \texttt{\textbf{DYN (User-driven)}} & \texttt{\textbf{DYN (Data-driven)}} &  & \texttt{\textbf{Dependent}} & \texttt{\textbf{Invariant}} & & \texttt{\textbf{3D Image}}\\

    %
    Easy to use* & Custom shape control* &  Aids feature discovery* & & Responsive to user intent* & Stable across view changes* & & Higher depth perception*\\
    Easy to implement* & Good shape–feature fit* & Tigher shape-feature fit*  & & Intuitive for immersion & Ignores view awareness & & Latency risk\\
    Poor feature coverage & Manual interaction effort & Reduced user effort*  & & Unintended effect changes \\
     &  & Data-sensitive fragility \\
     &  & Opaque logic \\
    \end{tabular}


    \begin{tabular}{@{}p{0.15\linewidth}p{0.15\linewidth}p{0.15\linewidth}p{0.035\linewidth}p{0.15\linewidth}p{0.15\linewidth}p{0.01\linewidth}p{0.15\linewidth}@{}}
    \texttt{\textcolor{colorScope}{\textbf{\large{Effect Scope}}}} &  &   & & \multicolumn{2}{l}{{\texttt{\textcolor{colorPOS} {\textbf{\large{Position, Orientation \& Scale}}}}}} & \\
    \cline{1-3}\cline{5-6}
    \vspace{1em} & & \\[-1.5ex]

    \texttt{\textbf{Interior}} & \texttt{\textbf{Exterior}} & \texttt{\textbf{Separate View}}& & \texttt{\textbf{Interactive}} & \texttt{\textbf{Semi-automated}} & & \texttt{\textbf{Decal Image}}\\

    Focus-context separation* & Reveals peripheral patterns* & Supports cascading effects* & & Full control* & Data-driven targeting* & & Strong spatial anchoring*\\
    Localized effect* & Enables multi-foci effects* & Persistent lens history* & & Fosters exploration* & Aids precision* &  & Less lens-induced clutter*\\
    Hides broader links & Weaker lens-effect binding & Cross-view burden & & Demands spatial dexterity & Misleads if heuristics fail & & Fragile on uneven surfaces\\
        &   &   &   &  Generalization liability
    \end{tabular}

\caption{Design trade-offs across our design dimensions. Entries marked with an asterisk (*) denote benefits; others indicate pitfalls.}
\label{tab:summaryTradeOffs}
\end{table*}


\begin{figure}[t!]
    \centering
    \includegraphics[width=1\linewidth]{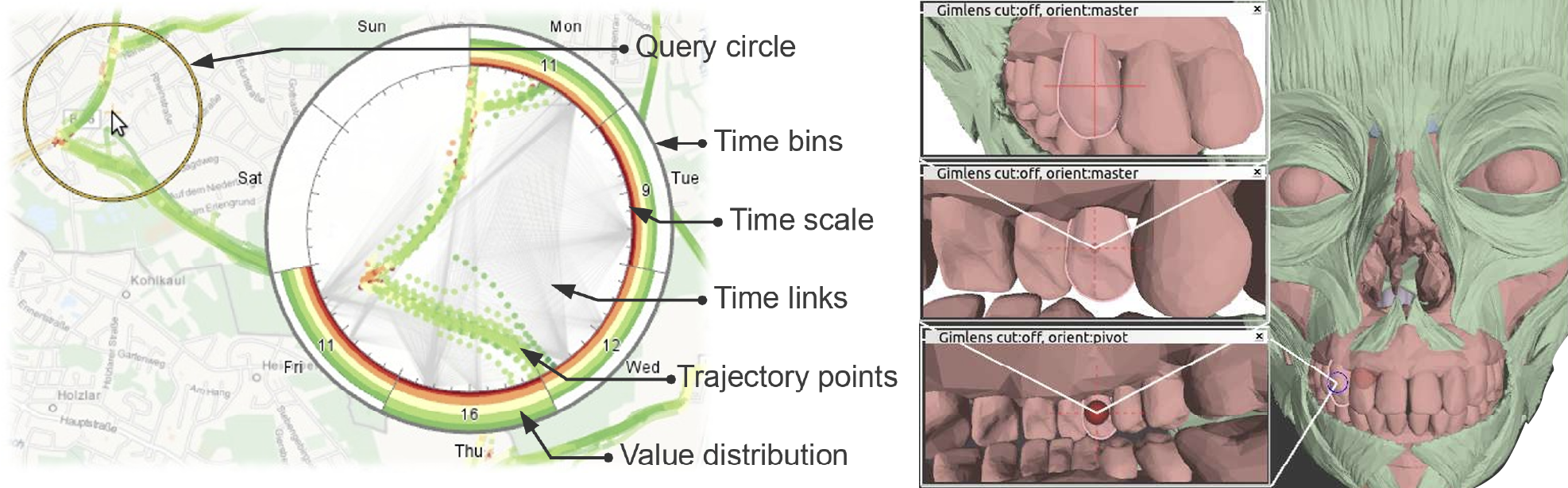}
    \vspace{-.75cm}
    \caption{ Lens displaying a \texttt{\textcolor{black}{\textbf{circular}}}, \texttt{\textcolor{colorEncoding}{\textbf{thematic lens image}}} of temporally aggregated trajectory information as a \texttt{\textcolor{black}{\textbf{separate view}}} \cite{tominski2012}(\textit{right}). Gimlens provides cascaded \texttt{\textcolor{colorEncoding}{\textbf{spatial images}}} in \texttt{\textcolor{black}{\textbf{independent views}}}, serving as complementary perspectives of a maxillary dental arcade \cite{pindat2013} (\textit{right}).} 
    \label{fig:exampleEncoding}
    \vspace{-.6cm}
\end{figure}

\vspace{-.08cm}
\subsection{\textcolor{black}{Effect Encoding}}
\vspace{-2px}

The lens effect enconding can be classified either as \texttt{\textcolor{colorEncoding}{\textbf{thematic}}} or \texttt{\textcolor{colorEncoding}{\textbf{spatially-based}}}, as shown in Fig. \ref{fig:exampleEncoding}. 

\texttt{\textcolor{colorEncoding}{\textbf{Thematic encoding}}} \iconEncodingThematic\ draws from information visualization and refers to abstract representations of the lens' data selection---\textit{e.g.} linear and radial histograms (Fig. \ref{fig:exampleEncoding}--\textit{left}) \cite{tominski2012, ma2020, liu2011visual}.
Hurter \textit{et al}. proposed a dual-layout lens that, within its boundary, rearranges pixels representing a 3D model's color-coded cells by transforming them from a Cartesian layout into a polar histogram layout \cite{hurter2011}.
Scheepens \textit{et al}. proposed a lens for map-based visualizations that generates \texttt{\textcolor{colorEncoding}{\textbf{thematic}}} images as \texttt{\textcolor{black}{\textbf{separate views}}} to show traffic flow over time \cite{scheepens2015}. Users can reposition and stack the views, triggering Boolean operations---\textit{e.g}. computing the difference between the originating views.
Conversely, \texttt{\textcolor{colorEncoding}{\textbf{spatially-based encoding}}} \iconEncodingSpatial\ draws from scientific visualization and concerns physically-based data representations---\textit{e.g.} glyphs, contours, and isosurfaces (Fig. \ref{fig:exampleEncoding}--\textit{right}) \cite{rocha2018, tong2016, gasteiger2011}.
Gasteiger \textit{et al.} presented a lens to support cerebral aneurysm analysis, where analysts must correlate surface attributes with and adjacent blood flow---\textit{e.g.} the wall shear stress and underlying inflow jet \cite{gasteiger2011}. This technique, places seed points within the lens boundary, generating illustrative color-coded streamlines, while surface attributes are depicted as isosurfaces or saturation-coded contour lines.
Rocha \textit{et al.} presented a \texttt{\textcolor{colorImagery}{\textbf{decal}}} lens to allows users to locally switch attributes on arbitrary surfaces, obtaining alternative data representations---\textit{e.g.} glyphs \cite{rocha2018}.

While both \texttt{\textcolor{colorEncoding}{\textbf{thematic}}} and \texttt{\textcolor{colorEncoding}{\textbf{spatial}}} mappings aim to represent data, they are shaped by different driving forces and usage purposes \cite{fairbairn2001representation}. 
\texttt{\textcolor{colorEncoding}{\textbf{Thematic}}} mapping emphasizes higher-level reasoning. It has been used to analyze multiple attributes simultaneously, and to uncover patterns not directly visible in spatial data---\textit{e.g.} correlations, trends, outliers, or temporal cycles \cite{pena2019comparison, sun2016embedding, andrienko2005visual}.
However, abstraction may introduce cognitive load by obscuring the underlying spatial structure of the data, requiring users to mentally relate abstract visuals back to the physical space.
In contrast, \texttt{\textcolor{colorEncoding}{\textbf{spatial}}} mapping emphasizes spatial reasoning by preserving the data's geometric and topological structure. This increases users' spatial awareness, facilitating the correlation of attribute data with physical referents. However, \texttt{\textcolor{colorEncoding}{\textbf{spatial}}} encodings---\textit{e.g.} contours, glyphs, or color mapping, heavily rely on users to interpret visual representations. With less abstraction to guide and summarize meaning, users themselves must perceive shapes, spatial relations, patterns, and other visual cues.
They are also prone to clutter, which reduces readability---particularly in dense or 3D environments. Thus, careful design is required to balance spatial fidelity with visual clarity.


\begin{figure}[t!]
    \centering
    
    \includegraphics[width=1\linewidth]{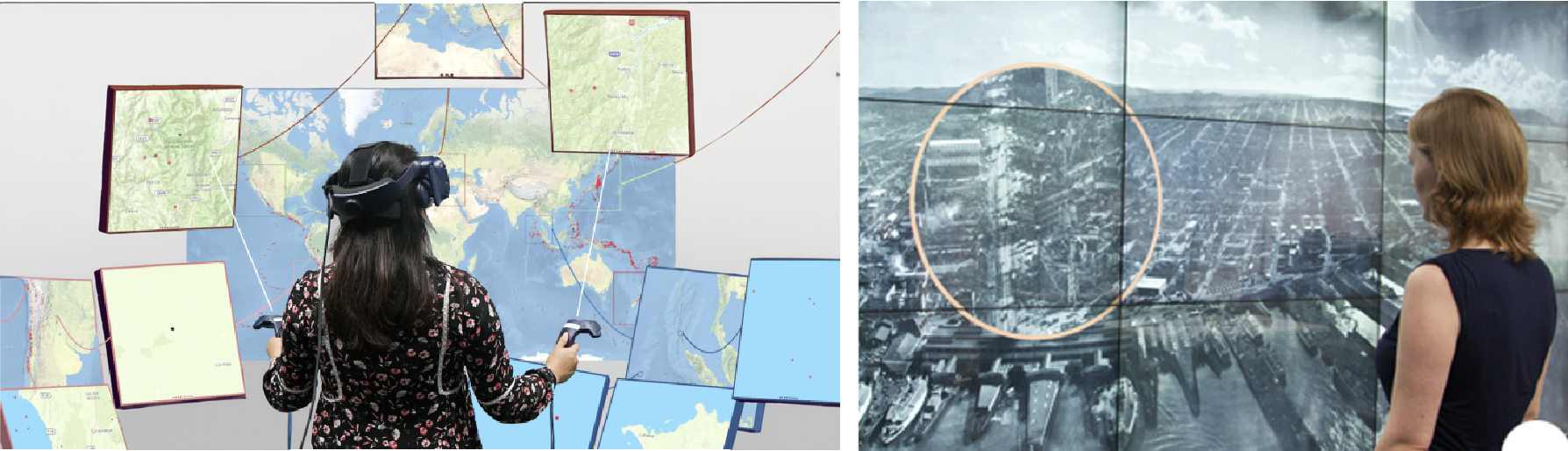}
    \vspace{-.7cm}
    \caption{ A hierarchical layout of magnified lens images, rendered as \texttt{\textcolor{black}{\textbf{separate views}}}, that remain \texttt{\textcolor{colorView}{\textbf{invariant}}} to the observer's viewpoint \cite{satriadi2020} (\textit{left}). 
    A \texttt{\textcolor{colorView}{\textbf{view-dependent}}}, \texttt{\textcolor{black}{\textbf{\textit{2D}}}} \texttt{\textcolor{black}{\textbf{circular}}} lens compensates for user's distance by automatically enlarging its content, ensuring that the user perceives it to be always the same size \cite{kister2015} (\textit{right}).} 
    \label{fig:exampleView}
    \vspace{-.6cm}
\end{figure}

\vspace{-3px}
\subsection{\textcolor{black}{Viewpoint Dependency}}
\vspace{-2px}

Lenses can be classified either as \texttt{\textcolor{colorView}{\textbf{dependent}}} or \texttt{\textcolor{colorView}{\textbf{invariant}}} from users' viewpoint, as shown in Fig. \ref{fig:exampleView}. This dependence can involve either the lens selection, function, effect, or even a combination thereof.

The degree to which a visualization depends on the user’s viewpoint is a key design consideration---especially in immersive spaces,where users can move physically \cite{reipschlager2020personal, jakobsen2013information}. This is particularly relevant for immersive focus+context visualization, where the focus reflects user's current interest.
However, physical locomotion can raise visibility and perceptual issues in the focused visualization: visual encodings may become imperceptible with increasing distance (leading to change blindness), distort under acute viewing angles, or even obstruct other foci distributed in the space.

Although most papers do not consider viewpoint in their designs, \texttt{\textcolor{colorView}{\textbf{view-sensitive}}} \iconViewDependent\ lenses afford fluid and intuitive interactions by dynamically adapting to users' physical navigation or gestures---\textit{e.g.} automatically expanding when users move away, and revealing fine-grained information when in close proximity (Fig. \ref{fig:exampleView}--\textit{right}) \cite{kister2015}. These dynamic responses allow lenses to better anticipate and align with user intent, likely reducing the need for manual input. However, design considerations are needed to avoid unstable and disorienting effects caused by unintentional movements---\textit{e.g.} to use smoothing or ``lock-in'' mechanisms.
For instance, a handheld mirroring lens should account for minor involuntary hand movements, common in immersive environments, to prevent jitter in the reflective effect \cite{bezawada2011investigation, bichlmeier2009, teather2009effects}.
%
%
%

Nearly \textit{all} lenses are viewpoint \texttt{\textcolor{colorView}{\textbf{independent}}}, ensuring stable lens images across view changes.
However, this comes at the cost of ignoring view awareness---\textit{e.g.} lenses may appear occluded, misaligned, or illegible from certain distances or angles. As a result, users must compensate through manual input, increasing interaction effort and possibly disrupting the analysis workflow---see Fig. \ref{fig:exampleView}--\textit{left}.
Ebert \textit{et al.} introduced tangible ring objects that function as lenses placed on tabletop displays, where users physically manipulate them to update their associated positions \cite{ebert2013tangiblerings}.
K\"{a}ser \textit{et al.} proposed magnifying lenses, where users control both position and scale with their fingers ~\cite{kaser2011fingerglass}.


\begin{figure*}[t!]
    \centering
    \includegraphics[width=1\linewidth]{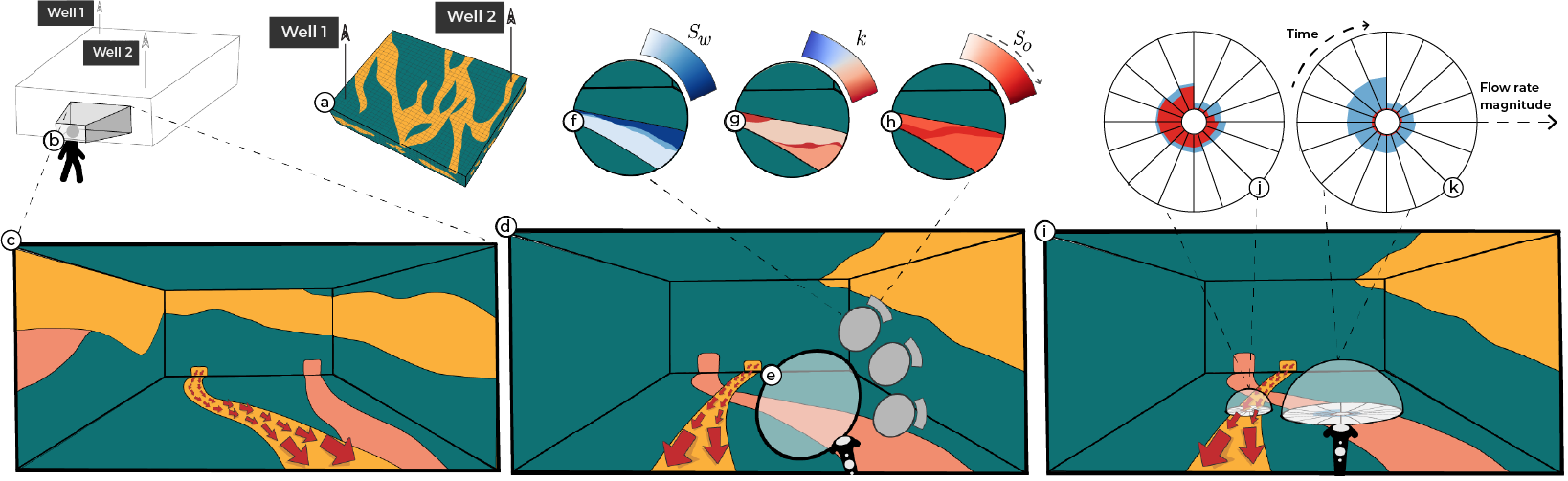}
    \vspace{-.8cm}
    \caption{Illustrative waterflooding production scenario, where water is injected into well $W_1$ to displace and recover reservoir oil at producer well $W_2$ (\textit{a}). Drawings of three composite lenses:  cutaway lens \textit{(b, c}), multiview
lens (\textit{d, e, f, g, h}), and thematic lens (\textit{i, j, k}).} 
    \label{fig:useCase2}
    \vspace{-.6cm}
\end{figure*}

\vspace{-3px}
\section{Generative Use Case}
\label{sec:caseStudies}
\vspace{-2px}

This section demonstrates the descriptive and generative utility of our design space by illustrating a synthetic case analysis that characterizes three spatially-embedded visualization lenses as functions of our dimensions.
This use case was motivated by the observed scarcity of research on composite lenses. \textit{Composite} lenses recursively combines multiple lens effects, enabling users to activate and deactivate individual effects as necessary. Additionally, our ideations for this case stems from our long-term collaboration with domain experts as well as prior experience in developing visualizations applied to reservoir engineering.

\vspace{-3.2px}
\subsection{Design Case: Composite Lenses}
\vspace{-2px}
\noindent In the oil and gas domain, a key visual analysis task is correlating variations in the flow behavior with attribute changes occurring in intertwined geological channels. 
To illustrate, one may imagine a scenario where water is injected from an injecting well $W_1$ to displace and produce oil in the producing well $W_2$, as depicted in Fig. \ref{fig:useCase2}--\textit{a}.

\vspace{3px}
To address the inherent occlusion caused by the volumetric nature of reservoir grids, an engineer activates an \texttt{\textcolor{black}{\textbf{interactive}}} \textit{cutaway lens}. This \texttt{\textcolor{black}{\textbf{3D}}} lens has the form of a \texttt{\textcolor{black}{\textbf{frustum}}} centered at the user’s point of view---see Fig. \ref{fig:useCase2}--\textit{b}, and continuously clips all reservoir grid cells located within its \texttt{\textcolor{black}{\textbf{interior}}}.  This clipping behavior produces a \texttt{\textcolor{black}{\textbf{projected}}} \texttt{\textcolor{black}{\textbf{cutaway}}} view  of the reservoir that dynamically changes according to the user's head and whole-body movements, thereby freeing their hands to perform other analysis tasks---\textit{e.g.} manipulating other lenses.
In this way, the \texttt{\textcolor{black}{\textbf{view-dependent}}} cutaway lens enables the engineer to physically navigate and inspect the internal parts of the reservoir grid---\textit{i.e.} the geo channels and water/oil flow paths.

\vspace{3px}
\noindent Figure \ref{fig:useCase2}--\textit{c} shows a cut region with two color-coded geo channels penetrated by the producer well $W_2$---yellow and orange channels, and the underlying oil flow represented by red arrow glyphs.
Near the well $W_2$, the engineer observes a steady oil flow coming from the leftmost yellow channel, while the rightmost orange channel produces no oil.
The engineer then proceeds along the orange channel towards the injecting well $W_1$, examining this unexpected absence of flow. 
Eventually, the engineer reaches a
location where the pathways intersect once again, with flow behavior akin to
that near $W_2$ (Fig. \ref{fig:useCase2}--\textit{d}).

To examine the local flow, the engineer activates an \texttt{\textcolor{black}{\textbf{interactive}}}, \texttt{\textcolor{black}{\textbf{view-invariant}}} \textit{multiview lens} (Fig. \ref{fig:useCase2}--\textit{e}). This \texttt{\textcolor{black}{\textbf{circular}}} \texttt{\textcolor{black}{\textbf{2.5D}}} lens captures the current viewpoint within its area.
It generates one or more  \texttt{\textcolor{black}{\textbf{color-coded}}}, \texttt{\textcolor{black}{\textbf{individual views}}}---\texttt{\textcolor{black}{\textbf{projected}}} images of the same region showing different attributes that influence flow, like water saturation $S_w$, permeability $k$, and oil saturation $S_o$ (Fig. \ref{fig:useCase2}--\textit{f, g, h}).
By inspecting the multi-attribute lens images, the engineer observes high permeability $k$ and oil saturation $S_o$ in the orange channel, but also a thief zone with elevated water saturation $S_w$. The engineer then suspects that the higher viscosity of oil compared to water causes the injected water from $W_1$ to preferentially flow through the thief zone, displacing only the lower-viscosity water and leaving oil behind.

\vspace{3px}
To confirm this hypothesis, the engineer activates an \texttt{\textcolor{black}{\textbf{interactive}}}, \texttt{\textcolor{black}{\textbf{view-agnostic}}} \textit{thematic lens} to inspect flow through two regions of interest within each geo channel. A \texttt{\textcolor{black}{\textbf{spherical}}} \texttt{\textcolor{black}{\textbf{3D}}} lens selects the reservoir grid cells contained in each region, generating a \texttt{\textcolor{black}{\textbf{thematic}}} view shown \texttt{\textcolor{black}{\textbf{inside}}} a \texttt{\textcolor{black}{\textbf{decal}}} lens image.
In regions with intricate geometry---\textit{e.g.} sharp discontinuities or pronounced curvatures, a redesign might use proxy surfaces as simplified, smoother approximations for decal placement \cite{rocha2018}.
This \texttt{\textcolor{black}{\textbf{decal}}}-based design reduces clutter and strengthens the spatial correlation between the lens imagery and the reservoir region in focus (Fig. \ref{fig:useCase2}--\textit{i}). The \texttt{\textcolor{black}{\textbf{thematic}}} view portrays a radial histogram of two temporal attributes for the selected grid cells: water (blue) and oil (red) flow rate variations ($r$-axis) over time ($\theta$-axis).

\vspace{3px}
By inspecting the two lens images, the engineer observes that the leftmost \texttt{\textcolor{black}{\textbf{thematic}}} view depicts the expected flow behavior---\textit{i.e.} similar rates of oil and water flow, as the injected water effectively sweeps oil from the yellow channel (Fig. \ref{fig:useCase2}--\textit{j}). Conversely, the rightmost \texttt{\textcolor{black}{\textbf{thematic}}} image indicates increasing water flow rates alongside negligible oil flow, confirming the hypothesis that these anomalous flow behaviors result from the formation of the prior thief zone for the injected water throughout the orange channel (Fig. \ref{fig:useCase2}--\textit{k}).

\vspace{-3px}
\section{Discussion, Research Directions \& Limitations}
\label{sec:discussionLimitations}
\vspace{-2px}
This section discusses open research directions for lens visualizations by examining underexplored areas in the design space---see Fig. \textcolor{black}{\ref{fig:codingDesignSpace}}.


\vspace{3px}
\noindent{{\textbf{Position, Orientation \& Scale}}} Our survey indicated that most lens techniques adopt \texttt{\textcolor{colorPOS}{\textbf{full interactivity}}} ($35$ out of $45$ papers), thus relying entirely on users to manipulate the lens parameters. Beyond that, nearly all \texttt{\textcolor{colorPOS}{\textbf{semi-automated}}} techniques consider conventional 2D spaces rather than 3D immersive environments ($9$ out of $10$).
Therefore, further research on \texttt{\textcolor{colorPOS}{\textbf{assisted}}} POS could be particularly beneficial in immersive mediums, not only for guiding lens placement toward interesting data features but also for assisting in precise lens manipulation within 3D spaces---since prior work provides evidence that spatial interactions often suffer from precision issues due to the absence of physical surfaces that provide force feedback \cite{arora2017experimental}.

Furthermore, \texttt{\textcolor{colorPOS}{\textbf{semi-automation}}} could benefit immersive lenses with cascaded  \texttt{\textcolor{black}{\textbf{images}}} displayed across various \texttt{\textcolor{black}{\textbf{separate views}}} by determining optimal layouts that assist logical linkage and sensemaking of information through spatial arrangements. As limited research has examined layout possibilities, open opportunities remain to explore novel \texttt{\textcolor{colorPOS}{\textbf{semi-automated}}} 3D spatial layouts and their trade-offs in immersive spaces, as they differ significantly from flat grid layouts used in conventional displays. To illustrate, a recent study found that users prefer 3D spherical cap layouts when assessing hierarchical lens-like images of maps with varied magnification levels in immersive spaces \cite{satriadi2020}. However, the study also revealed that users frequently rearrange the surrounding images during analysis tasks.


\vspace{3px}
\noindent{{\textbf{Shape}}} The shape is a key factor in determining  proper lens selection and its associated effect. As most lenses proposed in the past years are of \texttt{\textcolor{colorShape}{\textbf{fixed}}} shape ($34$ out of $45$ surveyed papers), users may have adapted to them and, as a result, must mentally compensate for their limitations. For instance, magnification lenses of \texttt{\textcolor{colorShape}{\textbf{static}}} shapes may fail to provide clear and unambiguous magnifications of irregular regions of interest---either the lens is too small, requiring continuous viewpoint adjustment as the region does not fully fit in the focus, or it is too big, encompassing the entire focus region but at the expense of also including surrounding context. In the latter case, users must mentally segment the presented magnified imagery; however, mental burden may become overwhelming when the focus data features are sufficiently intricate, and when motion parallax is missing as seen in static \texttt{\textcolor{black}{\textbf{2D projection}}} lens images.

As our survey revealed that lens shapes \texttt{\textcolor{colorShape}{\textbf{dynamically generated either by the user}}} or \texttt{\textcolor{colorShape}{\textbf{by the underlying data}}} is a widely underexplored area ($3$ papers), future research could investigate designing lenses whose shape dynamically adapts to the geometry or data of interest, while ensuring proper continuity when transitioning from the focus to the context areas. In this way, these conformal lens shapes would possibly disambiguate the local features in focus, considerably preserve the context, and ultimately reduce cognitive confusion for users when interpreting the lens results.


\vspace{3px}
\noindent{{\textbf{Effect Imagery}}} The vast majority of the surveyed lenses employ \texttt{\textcolor{colorImagery}{\textbf{2D projection}}} images ($42$ out of $45$ papers), followed by very few that use \texttt{\textcolor{colorImagery}{\textbf{3D}}} ($3$) and \texttt{\textcolor{colorImagery}{\textbf{decal}}} ($2$) images. The same pattern holds when considering only immersive environments.

Regarding \texttt{\textcolor{colorImagery}{\textbf{3D}}} lens images, as previously mentioned, they may resemble world-in-miniatures whose core affordance lies in serving as both an \textit{alternative} \textit{visualization} and \textit{interaction space} that supplement a large-scale virtual environment---\textit{e.g}. by selecting an object that is far away from the user’s location. 
Nonetheless, our survey revealed that volumetric images typically apply the same visual representations as the base visualization. However, there are likely possibilities for further work to investigate novel \texttt{\textcolor{colorImagery}{\textbf{volumetric}}} lens images that alter the focused visualization in order to depict alternative data facets or visual abstractions---\textit{e.g.} a lens could apply distortions or perspective cut-offs in its miniature image in order to enable visual access to its partially-occluded focus regions.
Beyond that, the world-in-miniature metaphor could be translated as lenses supplementing the base visualization by enabling 3D interactions within the lens imagery itself to adjust lens parameters.
Further research could investigate \texttt{\textcolor{colorImagery}{\textbf{3D}}} images with distinctive visual abstraction or interaction idioms.
\setlength{\columnsep}{5.5pt}
\begin{wrapfigure}{l}[0pt]{0.33\textwidth}  
  \centering
  \vspace{-.4cm}
  \includegraphics[width=0.32\textwidth]{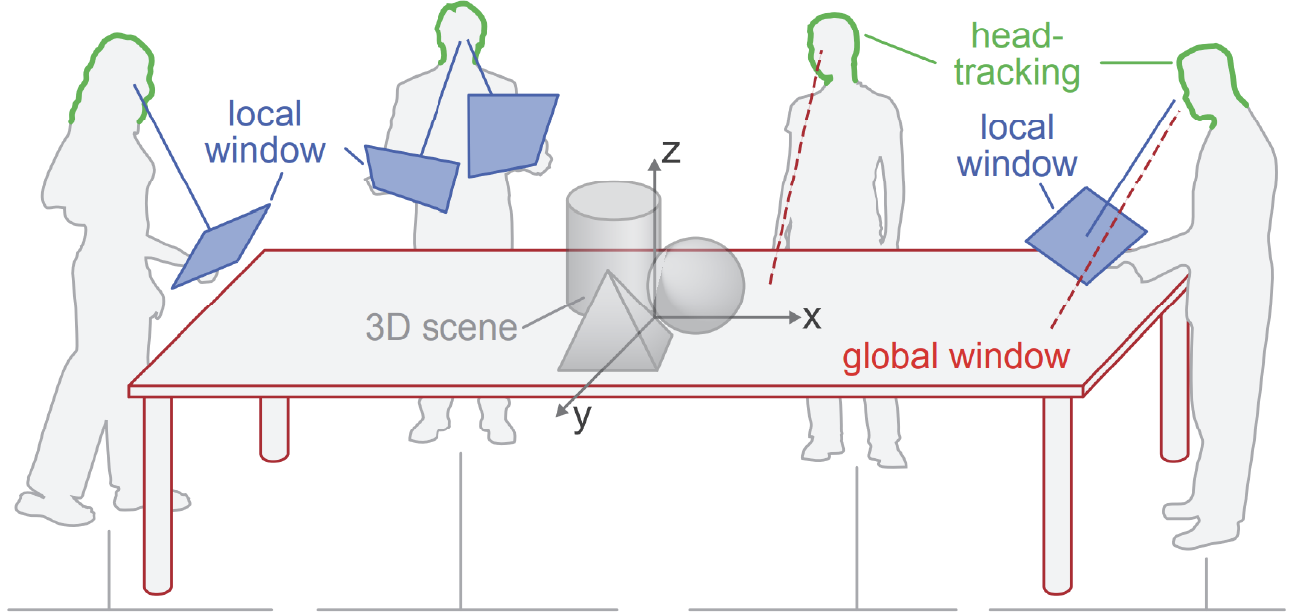}
  \vspace{-.4cm}
  \caption{Large collaborative virtual space serves as a \textit{global} view, while multiple tracked paper-based displays act as \textit{local} lens views \cite{spindler2012}.}
  \label{fig:sharedLens}
  \vspace{-.5cm}
\end{wrapfigure}
\noindent Furthermore, our survey revealed that very little research has investigated combining different categories of lens images ($2$ out of $45$ papers). This approach could be particularly beneficial for 3D visualizations, as they often contain a combination of multiple geometric data representations---\textit{e.g.} points, streamlines, and surfaces, each with associated attributes. In multi-geometry visualizations, users typically need to comprehend (interrelations among) these different spatial data structures in order to gain a holist understanding of the phenomena. 
In this sense, further research could investigate the design and added benefits of merging different types of lens images, applying each to the data type they are best suited for---\textit{e.g.} \texttt{\textcolor{colorImagery}{\textbf{3D}}} images are suitable for point clouds, while \texttt{\textcolor{colorImagery}{\textbf{decal}}}-based images best operate on surfaces.


\vspace{3px}
\noindent{{\textbf{Viewpoint Dependency}}}  \texttt{\textcolor{colorView}{\textbf{Viewpoint responsiveness}}} is a heavily underrepresented design strategy, with only $4$ papers in our survey featuring it---this holds true in immersive environments. 
On conventional displays, viewpoint dependency may be overlooked since the limited information space occupies users' full attention---thus, focus only shifts with explicit interaction. 
In contrast, on large immersive space, viewpoint changes may signal shifts in user attention and perception of the focus (and context) within the visualization. Further research could consider novel \texttt{\textcolor{colorView}{\textbf{view-sensitive}}} lens designs that adapt their emphasis, effect, function, or even a combination thereof.

Moreover, our survey revealed that immersive lenses are typically placed in close, personal spaces directly in front of users \cite{spindler2012, kister2015, mota2018}. This is intuitive as the focus is often something that is perceived to be near the viewer. However, in large virtual environments, this may not always hold true---\textit{i.e.} when users virtually or physically move, the focus of the visualization might still match the center of their attention. To illustrate, in large 3D urban models, continuous perspective changes poses a significant challenge for urban planners when comparing points of interest within the city---\textit{e.g.} buildings, that are not only distant from each other but may also be occluded by surrounding buildings, depending on the user's viewpoint \cite{chen2021urbanrama}. Therefore, future research could investigate design changes in immersive lenses situated outside users' personal space that still reflect their focus of information.

Furthermore, our survey indicated that the few \texttt{\textcolor{colorView}{\textbf{view-dependent}}} immersive lenses were mainly focused on supplementing \textit{individual} work, even in \textit{collaborative} environments---see Figure \ref{fig:sharedLens}. 
Therefore, future research could tackle design aspects on \texttt{\textcolor{colorView}{\textbf{perspective-sensitive}}} lenses to support shared work, where multiple users may look at lens images from different points of view. 
For instance, one could imagine a user communicating focused information to a co-worker situated elsewhere, who has a partially occluded or a relatively narrow viewing angle of the focus region. 
This could involve rearranging the images to increase its visual angle and perception for the second viewer.


\vspace{1px}
\noindent{{\textbf{Limitations}}} Lastly, we want to acknowledge that this paper is not a comprehensive literature survey---nor was it intended to be. While our sampled corpus spans work from the past 15 years to reflect a degree of diversity, limitations in sample size and representativeness remain, as the corpus construction mainly relied on manual search and annotation. For that, we refer to an extensive survey conducted by Tominski \textit{et al}. on lenses at a general level \cite{tominski2017}.

\nocite{krueger2014visual, lu2017visual, lu2015trajrank, coffey2011, van2010, butkiewicz2008multi, tong2015view,looser20073d, elmqvist2007occlusion, zhang2020clusterlens, niebling2017spatial, schott2020virtual, coffin2006interactive}


\vspace{-7px}
\section{Conclusion}
\vspace{-2px}
In this work, we tackled the question of \textit{what are design components and choices to be considered when constructing spatially-embedded visualization lenses?} 
To this end, we conducted a survey and collected a corpus of 45 papers published in the visualization literature over the past 15 years. 
From this corpus, we derived a design space comprising seven design dimensions. For each dimension, we distilled information on specific design choices---their usage scenarios, advantages, and disadvantages, supported by a variety of examples. 
Following this, we described a detailed generative use case encompassing three immersive lenses, thereby conveying the expressive power of our design model.

Finally, guided by our design space, we identified and discussed opportunities for future research on lens design---considering both conventional and emerging immersive environments.  For designers, our catalog surfaces a range of design archetypes along with their trade-offs.
For researchers, we hope that our work not only stimulates investigations on new designs but also provides a shared vocabulary for how future research understands, describes, and discusses the value of spatially-embedded lens visualizations.


\bibliographystyle{abbrv-doi}

\bibliography{template}
\end{document}